\documentclass[%prl,
aps,%preprint %tightenlines
 ,twocolumn
]{revtex4}
\usepackage{graphicx}

\begin{document}

\title{Entanglement decoherence in a gravitational well according to the event formalism
}

\author{ T.C.Ralph$^1$ and J.Pienaar$^2$}\affiliation{%Centre for Quantum Computation and Communication Technology, \\
$^1$School of Mathematics and Physics, University
of Queensland, Brisbane, Queensland 4072, Australia \\
$^2$ Faculty of Physics, University of Vienna, Boltzmanngasse 5, 1090 Vienna, Austria}

\date{\today}

\begin{abstract}
{The event formalism is a non-linear extension of quantum field theory designed to be compatible with the closed time-like curves that appear in general relativity. Whilst reducing to standard quantum field theory in flat space-time the formalism leads to testably different predictions for entanglement distribution in curved space. In this paper we introduce a more general version of the formalism and use it to analyse the practicality of an experimental test of its predictions in the earth's gravitational well.}

\end{abstract}

%
%\pacs{03.67.Dd, 42.50.Dv, 89.70.+c}

\maketitle

%\tableofcontents

\vspace{10 mm}
\section{Introduction}

A complete theory of quantum gravity has remained elusive for almost a century since the discovery of quantum mechanics. While quantum mechanics is well tested and confirmed on Earth's surface, where gravity is approximately uniform, there is little experimental data on quantum systems across significant gravitational potentials. This is true even for relatively accessible regimes such as the gravitational potential of Earth, for which there exist well-established theoretical models based on semi-classical techniques. Given the lack of data, and the lack of consensus on a fundamental theory of quantum gravity, there is still potential for new experimental discoveries to be made in these regimes.

In proposing such experiments, it is important to consider alternatives to the usual semi-classical approach \cite{BIR82}. On one hand, such alternative theories challenge the status-quo and encourage experimental tests to deepen our understanding of the existing paradigm. Less appreciated, however, is the role of such alternatives in hypothesis testing: they allow us to design effective experiments. In particular, a negative result for the standard hypothesis might provide positive support to an alternative theory, instead of being written off as anomalous or due to experimental error. The role of alternative theoretical models in driving and guiding experimental progress therefore should not be understated. 

At the interface between quantum mechanics and gravity at the meso- or macroscopic scale, most models predict a decoherence-like effect on quantum entanglement and quantum superpositions. However, the precise mechanism of decoherence and its relation to gravity tends to differ widely between approaches (see eg. Ref. \cite{AN3} for a more complete discussion). The more conservative approaches focus on weak gravitational fields in a semi-classical setting\cite{AN1,AN2,HU1,BLE,LAM}; in addition, decoherence due to centre-of-mass coupling to internal degrees of freedom in the presence of time dilation has also recently been considered\cite{PIK}. On the other hand, more radical objective state-reduction models call for a break-down of quantum mechanics\cite{PEN,DIO,KAF}. These latter models are related to a famous thought experiment of Penrose, in which it was argued that a massive object placed in a superposition should quickly decohere in the position basis due to the inherent uncertainty induced in the space-time metric. By contrast to the above approaches, the mechanism considered in the present work is based on a completely different thought experiment due to Deutsch: the self-consistent dynamics of quantum systems near closed time-like curves\cite{DEU}. 

Deutsch's thought experiment %, although also an example of a quantum gravity scenario, 
is perhaps less well-known than Penrose's. It considers exotic space-times in which gravity creates closed time-like curves and hence permits time-travel into the past. Deutsch argued that the usual paradoxes associated with such solutions of general relativity can be resolved by quantum mechanics. Deutsch does not attempt to quantise gravity, but considers quantum systems localised to semi-classical trajectories in a classical background space-time. Deutsch argues that a system scattering from a closed time-like curve in space-time exhibits globally non-linear and non-unitary dynamics. The {\it event formalism} extrapolates Deutsch's model to massless fields propagating in a globally hyperbolic space-time background, in which case it predicts a de-correlation of entanglement due to gravitational curvature \cite{RMD}. Unlike Penrose and other models that also treat space-time classically and posit a non-linear dynamical equation, the event formalism has a number of novel features: it predicts decoherence only for entangled systems and not single systems in a superposition; the effect is in principle reversible by further gravitational interactions (hence it is better called `de-correlation' than decoherence); and it may exhibit information processing power greater than that of standard quantum mechanics\cite{PIE}. 

Nonlinear modifications of quantum mechanics, whether due to gravity or otherwise, tend to be susceptible to pathologies and in particular to faster-than-light signalling, which is widely considered to be unphysical, particularly in a relativistic setting like quantum gravity. While the theoretical soundness of the Schr\"{o}dinger-Newton equation remains a topic of debate\cite{AN1,AN2,HU1}, recent work by Kent\cite{KEN05} indicates that it is possible to formulate a sub-class of non-linear theories that are manifestly free of such pathologies. So far, few physically motivated models have been constructed using this idea, but Deutsch's model is one example that can be cast in this form\cite{CAV12}, as is the model's extension to quantum optics in gravitational fields that we consider here. The present model is therefore theoretically interesting, as well as being experimentally testable.

In this paper we calculate the expected decoherence effect from distributing time energy entanglement from a ground station to a detector in orbit. The calculation is based mainly on the event formalism introduced by Ralph, Milburn and Downes in Ref. \cite{RMD}, but the formalism is generalized to more realistically account for the expected experimental situation. We begin by introducing the event formalism in a more general way. For simplicity we restrict ourselves to $1+1$ conformal space-times that admit foliation into space-like hyper-surfaces with respect to some global time parameter $t$. We use units in which $c=\hbar=1$.

\section{Event Formalism}

Deutsch considers a situation in which a particle (qubit) interacts with a future incarnation of itself via a closed timelike curve and shows this situation can be solved consistently\cite{DEU}. The event formalism makes minimalist modifications to quantum optics on a curved background such that it reproduces the predictions of Deutsch in appropriate limits for spacetimes containing closed timelike curves formed via wormhole type metrics \cite{RD12}. In Deutsch's model, operators defined at later times along the particles' trajectory commute with operators defined at earlier times; this allows the future version of a system to interact with itself in the past. Equivalently, we can think of the time-travelling particle as being represented by a pair of particles, one labelled `younger' and the other `older'. The younger particle disappears at time $t=t_F$ and the older particle appears at time $t=t_P$ (where $t_P < t_F$) and the initial state of the older particle is required to be equal to the final state of the younger particle. We can define a parameter $\tau$ that is monotonically increasing along the semi-classical trajectory of the single time-travelling particle and we can associate two Hilbert spaces to the same particle: one for its younger self $\tau<t_F$ and the other for its older self $\tau>t_P$. At times $t_P<t<t_F$ for which the parameter $\tau$ is two-valued, the Hilbert space of the particle is doubled, allowing the past and future versions to interact. The result is a non-linear map from the state just before $t_P$ to the state just after $t_F$.

The generalisation of the model to fields follows similar reasoning: we replace the point particle by a wavepacket, centred %at all times 
on a well-defined semi-classical trajectory. The parameter $\tau(t,t_d)$ is a monotonically increasing function of the global time $t$, and is defined for all times up to the detection of the propagating mode at time $t_d$. Specifically, $\tau$ is the time elapsed from $t$ until $t_d$, as measured incrementally by a set of local observers, all at rest with respect to the choice of co-ordinates $(x, t)$ and stationed along the semi-classical trajectory of the wavepacket. %As we will see below, 
Physical quantities will not depend directly on $\tau$ but rather on the %Lorentz-invariant quantity $\Delta$, which measures the 
relative mismatch in this parameter induced between two different modes of the field by propagation along different paths in space-time. Using this definition of $\tau$, we require that operators acting on the field at sufficiently different values of $\tau$ should commute with each other. For this purpose we introduce $\Omega$, the Fourier complement of $\tau$, and modify the standard commutation relations as follows. Given an appropriate choice of coordinates the standard quantum optical mode annihilation operator can be written
\begin{eqnarray}
%\hat a(z) &=& \int d \vec k L(k_x, k_y) K(k_z) e^{-i k_z z} \hat a_{\vec k}
\hat a_K(x, t) &=& \int d k K(k) e^{i k (x-t)} \hat a_{k}.
\label{e1}
\end{eqnarray}
The event formalism generalizes this to the event operator
\begin{eqnarray}
\bar a_K(x, t, \Delta) &=& \int d k K(k) e^{i k (x-t)} \int d\Omega |K(\Omega)| e^{i \Delta \Omega} \hat a_{k,\Omega} \nonumber\\
%\bar a(z, \Delta) &=& \int dk L(k_x, k_y) K(k_z) e^{-i k_z z} \int d\Omega |K(\Omega)| e^{i \Delta \Omega} \hat a_{\vec k, \Omega}
\label{e2}
\end{eqnarray}
The operators $\hat a_{k, \Omega}$ behave as standard boson operator with the commutator
\begin{eqnarray}
[\hat a_{k, \Omega}, \hat a_{k', \Omega'}^{\dagger}] = \delta(k - k') \delta(\Omega - \Omega')
\label{e3}
\end{eqnarray}
and the property that they annihilate the vacuum: $\hat a_{k, \Omega} |0 \rangle = 0$. The same-time event commutator is defined
\begin{eqnarray}
[\bar a_K(x, t, \Delta), \bar a_{K'}^{\dagger}(x', t, \Delta')]_e = {{[\bar a_K(x, t, \Delta), \bar a_{K'}^{\dagger}(x', t, \Delta')]}\over{\int d\Omega |K(\Omega)K'(\Omega)|}} \nonumber\\ 
\label{e4}
\end{eqnarray}
where the normalization term is necessary to avoid double counting the mode overlap. Eq.\ref{e4} has the following properties:
\begin{eqnarray}
&& \textrm{If} \;\;\;\;  \Delta - \Delta ' = 0  \nonumber\\
&& [\bar a_K(x, t, \Delta), \bar a_{K'}^{\dagger}(x', t, \Delta')]_e  =  [\hat a_K(x, t), \hat a_{K'}^{\dagger}(x', t)] \nonumber\\
\label{e5}
\end{eqnarray}
and
\begin{eqnarray}
&& \textrm{If}  \;\;\;\;  \Delta - \Delta ' \ne 0  \nonumber\\
&& |[\bar a_K(x, t, \Delta), \bar a_{K'}^{\dagger}(x', t, \Delta')]_e|  <  |[\hat a_K(x, t), \hat a_{K'}^{\dagger}(x', t)]| \nonumber\\
\label{e6}
\end{eqnarray}
Eq.\ref{e5} guarantees that when $\Delta - \Delta ' = 0$ all commutators reduce to those predicted by the mode operators. Eq.\ref{e6} guarantees that when $\Delta - \Delta ' \ne 0$ Eq.\ref{e4} is still a well behaved commutator. $\Delta$ parametrizes the difference between the globally defined detection time $t_d$ and a locally defined time $\tau(t)$:
\begin{eqnarray}
\Delta &=& t_{d}-\tau(t, t_d).
\label{del2}
\end{eqnarray}
As noted earlier, the parameter $\tau(t, t_d)$ records the propagation time between the detection time, $t_d$, and $t$, as incrementally measured by a set of local observers along the light path of this particular mode, i.e. 
\begin{equation}
\tau(t, t_d) =  \int^{t_d}_{t} ds
\label{ans}
\end{equation}
where $ds$ is the propagation time across an incremental local frame. We require that these local frames are all at rest with respect to the chosen frame of reference, i.e. with respect to the choice of co-ordinates $(x, t)$. This co-ordinate dependence of $\Delta$ is necessary to ensure that all physical predictions are reference frame independent (see end of this section).

%the total expression for the event operator Eq.\ref{e2} is Lorentz invariant; in particular, a change of reference frame will result in a change in $\Delta$ to compensate for the change in the complementary variable $\Omega$. (This is completely analogous to how Lorentz invariance is enforced in the standard mode operator Eq.\ref{e2}, where the contraction in length is perfectly compensated by the corresponding dilation of wavenumber in the exponent $k (x-t)$, under a change of reference frame.) 

For a sufficiently large space-time curvature as measured by $\Delta$, the operators commute at different times along the trajectory. If the system traverses a closed time-like curve, this ensures that the Deutsch model is recovered\cite{PIE,PIE2}. Conversely, for an inertial detection frame in flat space, all the local observers along the mode paths are in the same inertial frame (for example, the detection frame) so from Eq.\ref{ans}, $\tau = t_d - t$. Hence for this situation we have $\Delta = t$ and hence $\Delta - \Delta' = 0$ in all same-time commutators and we recover the standard theory. For curved space in general (not necessarily containing closed time-like curves), $\Delta \ne t$ and hence for modes that follow different paths we can have $\Delta - \Delta' \ne 0$, potentially leading to non-linear effects.

These definitions are sufficient to write down a simple recipe for calculating expectation values in the Heisenberg picture with the event formalism. First, write the desired expectation value in terms of a Hermitian function of mode operators representing the final measurement, $M(\hat a_K, \hat a_K^{\dagger},...)$, acting on an initial state formed via a unitary transformation of the global ground state $|\phi \rangle = U(\hat a_{K'}, \hat a_{K'}^{\dagger},...)|0 \rangle$. The distinction between $K$ and $K'$ here represents the possibility that the measurement and preparation modes differ. We obtain
\begin{eqnarray}
\langle M \rangle &=& \langle 0|U^{\dagger} M U |0 \rangle \nonumber\\
&=& \langle 0| M' |0 \rangle_{t_i}.
\label{exp1}
\end{eqnarray}
where $M'(\hat a_K, \hat a_K^{\dagger},\hat a_{K'}, \hat a_{K'}^{\dagger},...) = U^{\dagger} M U$ is the Heisenberg picture measurement operator and the subscript $t_i$ indicates that all mode operators are evaluated at the same initial time ($t_i$). The equivalent event expectation value is obtained by simply replacing mode operators with event operators (and mode commutators with event commutators) in Eq.\ref{exp1} such that
\begin{eqnarray}
\langle M \rangle_e &=& \langle 0| \bar M' |0 \rangle_{t_i}
\label{exp2}
\end{eqnarray}
where $\bar M' = M'(\bar a_K, \bar a_K^{\dagger}, \bar a_{K'}, \bar a_{K'}^{\dagger},...)$. The definitions given in Eq.\ref{e2} - \ref{e4} are then sufficient to calculate the expectation value. Notice that in flat space the fact that $\Delta -\Delta' = 0$ means that all expectation values will be the same as their mode operator equivalents. 

Notice that expectation values only depend on the values of same-time commutators. In the standard same-time mode commutator (RHS of Eq.\ref{e5}) Lorentz invariance is ensured because a change of reference frame leads to $|K(k)|^2 e^{i k(x-x')} \to |K(k')|^2 e^{i k'(x-x')}$ in the new frame, provided a suitable transformation of the dummy variable $k \to k'$ is made \cite{note}. As a result integrals in the commutator remain invariant under the change of reference frame. Similarly, we also have $|K(\Omega)|^2 e^{i \Omega(\Delta-\Delta')} \to |K(\Omega')|^2 e^{i \Omega'(\Delta-\Delta')}$ under a change of frame.
% provided a suitable transformation of the dummy variable $\Omega \to \Omega'$ is made.
This in turn ensures that the same-time event commutator, Eq.\ref{e4}, and hence all expectation values in the event formalism, are reference frame independent.

\section{Experimental Proposal}

We consider a generalized version of the scenario analysed in \cite{RMD} in which time energy entanglement is produced by a down converter and distributed along two paths that experience different curvatures (see Fig.1). We will first consider a general scenario in which the detectors and sources can be placed at arbitrary heights within the gravitational field of the earth. We will then consider a specific, realistic scenario in which the source and one of the detectors is on the surface of the earth whilst the other detector is in low earth orbit.

The gravitational field of the earth is modelled via the Schwarszchild metric in polar coordinates \cite{BIR82}:
\begin{eqnarray}
ds^2& =& \left(1-\frac{2M}{r}\right)dt^2-\left(1-\frac{2M}{r}\right)^{-1}dr^2 \\ \nonumber
&& -r^2(d\theta^2+\sin^2\theta d\phi^2).
\label{M1}
\end{eqnarray}
An infinitesimal proper distance in space-time as measured by a local observer is denoted $ds$. The proper distance between two space-time points is an invariant quantity, which all observers can agree on, regardless of the coordinate system they use. Here $dt$ is an infinitesimal time interval as measured by an observer far from the earth, $r = \textrm{circumference}/2 \pi$ is the reduced circumference, and $\theta$ and $\phi$ are spherical coordinates that remain the same for the observer on earth and the far away observer . We are using units where $G$, the universal gravitational constant and $c$, the speed of light, are both set to unity. $M$ is the mass of the earth expressed in geometric units (metres). 

In particular we consider the production of time energy entanglement from vacuum inputs via a parametric unitary.
\begin{figure}[htb]
\begin{center}
\includegraphics*[width=9.5cm]{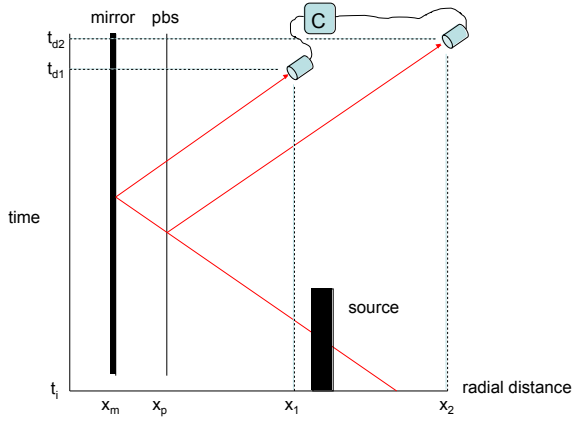}
\caption{Schematic of generic correlation experiment. The source populates a vacuum mode with photon pairs of orthogonal polarizations that propagate towards a massive body. The pairs are split up and reflected away from the body to two different photon counting detectors using a polarizing beamsplitter (pbs) and a mirror. The photo-currents from the photon counters are sent to a correlator (C) to count coincidences.}
\label{fig1}
\end{center}
\end{figure} 
%
%%
%\begin{eqnarray}
%\hat U^{\dagger}(\chi) \hat a_1\hat U(\chi) & = & Cosh\chi_{\textrm{max}}\; \hat a_1 + Sinh\chi_1\; \hat a_{2c}^{\dagger}\nonumber\\
%\hat U^{\dagger}(\chi) \hat a_2\hat U(\chi) & = & Cosh\chi_{\textrm{max}}\; \hat a_2 + Sinh\chi_2 \; \hat a_{1c}^{\dagger}
%\label{gg}
%\end{eqnarray}
%
Using the recipe outlined above and the properties of parametric amplification \cite{BAC03} we obtain the following event operators
\begin{eqnarray}
\bar a_{m1}' &=& \textrm{Cosh}(\chi_{\textrm{max}}) \bar a_{m1} + \textrm{Sinh}(\chi_1) \bar a_{m2c}^{\dagger} \nonumber\\
\bar a_{m2}' &=&  \textrm{Cosh}(\chi_{\textrm{max}}) \bar a_{m2} + \textrm{Sinh}(\chi_2)  \bar a_{m1c}^{\dagger}
\label{am18}
\end{eqnarray}
where
\begin{eqnarray}
\bar a_{m1} &=&  \int dk \; G(k) \;e^{ik(-x_{i1}-2 M ln(x_{i1}) - t_i+\phi_1^-)} \nonumber\\
 &\times& \;\;\; \int d\Omega \; |G(\Omega)| \;e^{i\Omega(t_{d1}-\tau_1(t_i))}  \bar a_{1,k,\Omega}\nonumber\\
 \bar a_{m2} &=&  \int dk \; G(k) \;e^{ik(-x_{i2}-2 M ln(x_{i2}) - t_i+\phi_2^-)} \nonumber\\
 &\times& \;\;\; \int d\Omega \; |G(\Omega)| \;e^{i\Omega(t_{d2}-\tau_2(t_i))}  \bar a_{2,k,\Omega}
\label{m12}
\end{eqnarray}
and
\begin{eqnarray}
\bar a_{m2c} & = &   \int dk \; H(k) \;e^{ik(-x_{i2}-2 M ln(x_{i2}) - t_i+\phi^c)} \nonumber\\
 &\times& \;\;\; \int d\Omega \; |H(\Omega)| \;e^{i\Omega(t_{d1}-\tau_1(t_i))}  \bar a_{2,k,\Omega}\nonumber\\
\bar a_{m1c} & = &  \int dk \; H(k) \;e^{ik(-x_{i1}-2 M ln(x_{i1}) - t_i+\phi^c)} \nonumber\\
 &\times& \;\;\; \int d\Omega \; |H(\Omega)| \;e^{i\Omega(t_{d2}-\tau_2(t_i))}  \bar a_{1,k,\Omega}\nonumber\\
\label{bars}
\end{eqnarray}
and
\begin{eqnarray}
\chi_j & = &  \int dk \; G_j(k) H(k)^{*} \;e^{ik(\phi_j^{-} -\phi^c)} \chi_{\textrm{max}}
\label{ggg}
\end{eqnarray}
with $j=1,2$.

The function $G(k)$ is the mode function of the detector, whilst $H(k)$ is the mode function of the source. In the general recipe of Eq.\ref{exp1} - \ref{exp2} these correspond to $K$ and $K'$ respectively.
%The event function $J_K(\Omega)$, $K=G,H$, characterizes the deviation from standard quantum mechanics on a curved background produced by a speculated decay in the commutator between operators lying on the same geodesic but having propagated according to differing local metrics. 
For simplicity we will consider the case of weak parametric amplification for which $\textrm{Cosh}(\chi) \approx 1$ and $\textrm{Sinh}(\chi) \approx \chi$. Under this condition and unit transmission and detection efficiency the rate of coincidence detection according to the event formalism is given by
\begin{eqnarray}
C & = &  \langle \bar{a}'^{\dagger}_{m1} \; \bar{a}_{m1}' \bar{a}'^{\dagger}_{m2} \; \bar {a}_{m2}'  \rangle \nonumber\\
& = &  \chi_2 \chi_1^{*}  [ \bar{a}_{m1},  \bar{a}_{m1c}^\dagger]_e \times [ \bar{a}_{m2},  \bar{a}_{m2c}^\dagger]_e ^{*} \nonumber\\
& = &  |\chi_{\textrm{max}}  \int \int dk dk' G(k) H(k)^{*} e^{ik(\phi_1^{-} -\phi^c)} \nonumber\\
&\times& %\;\;\;\;\;\;\; \;\;\;\;\;\;\; \;\;\;\;\;\;\; 
G(k')^{*} H(k') e^{-ik'(\phi_2^{-}-\phi^c)} {{\int d\Omega |G(\Omega) H(\Omega)| e^{i\Omega \Delta_t}}\over{ \int d\Omega |G(\Omega) H(\Omega)| }}|^2 \nonumber\\
\label{coin}
\end{eqnarray}
where 
\begin{eqnarray}
\Delta_t &=& \Delta_1-\Delta_2 \nonumber\\
&=& t_{d1}-\tau_1(t_i, t_{d1}) - (t_{d2}-\tau_2(t_i, t_{d2})).
\label{comm}
\end{eqnarray}
Notice that if $\Delta_t=0$ the $\Omega$ integral quotient goes to $1$. The expression for $C$ then reduces to its standard quantum optical prediction. Here we are assuming that $t_{d2} > t_{d1}$ and hence that the detectors are space-like separated. We will consider the case of time-like separated detectors in Section IV.

Assuming radial propagation of narrow (with respect to earth radius) beams gives \cite{RMD}
\begin{eqnarray}
\phi_1^- &=&  2 x_m + 4 M ln(x_m) + (t_{d1}-x_{d1}-2M ln(x_{d1})) \nonumber\\
\phi_2^- &=&  2 x_p + 4 M ln(x_p) + (t_{d2}-x_{d2}-2M ln(x_{d2})) \nonumber\\
\label{m13}
\end{eqnarray}
describing the phase shifts acquired by the modes through propagation. Here $x_m$, $x_p$, $x_{d1}$, $x_{d2}$, $t_{d1}$ and $t_{d2}$ are defined in Fig.1. The phase shifts acquired by the event degree of freedom are obtained by integrating the time coordinates along a series of stationary shell frames connecting the source to the detectors \cite{RMD}
\begin{eqnarray}
\tau_1 &=&  \int^{x_{d1}}_{x_m} {{d r'}\over{\sqrt{1-{{2 M}\over{r'}}}}}+ \int^{x_{i1}}_{x_m} {{d r'}\over{\sqrt{1-{{2 M}\over{r'}}}}}.\nonumber\\
& \approx & -t_i +t_{d1} - M ln({{x_{d1} x_{i1}}\over{x_m^2}})\nonumber\\
\tau_2 &=&  \int^{x_{d2}}_{x_p} {{d r'}\over{\sqrt{1-{{2 M}\over{r'}}}}}+ \int^{x_{i2}}_{x_p} {{d r'}\over{\sqrt{1-{{2 M}\over{r'}}}}}.\nonumber\\
& \approx & -t_i +t_{d2} - M ln({{x_{d2} x_{i2}}\over{x_p^2}}).
\label{tau2}
\end{eqnarray}
where we have simplified the results by assuming $r>>2 M$ for all radii of interest. Also $t_i$ is an arbitrary initial time, $x_{i1} = -t_i + 2 x_m + 4 M ln(x_m) + t_{d1}-x_{d1}-2M ln(x_{d1})$ and $x_{i2} = -t_i + 2 x_p + 4 M ln(x_p) + t_{d2}-x_{d2}-2M ln(x_{d2})$. This leads to
\begin{eqnarray}
\Delta_t &=& M ln({{x_{d1} x_{i1} x_p^2}\over{x_{d2} x_{i2} x_m^2}}).
%\nonumber\\
%& \approx & 2 M ln({{x_p}\over{x_m}})   
\label{comm}
\end{eqnarray}
%
%
%The event functions $J_K(\Omega)$ are not directly specified by the theory. However, they are expected to reflect the temporal quantum uncertainty in the process of producing and measuring the state and are constrained by the requirement that standard quantum mechanical predictions should emerge in the limit of inertial observers in flat space. We propose two possible forms for these functions:
%%
%\begin{eqnarray}
%J_K &=& G \;if \; |\int d\Omega |G(\Omega)|^2 e^{\Omega \Delta}| > |\int d\Omega |H(\Omega)|^2 e^{i\Omega \Delta}| \nonumber\\
%&=& H \; if \; |\int d\Omega |H(\Omega)|^2 e^{\Omega \Delta}| > |\int d\Omega |G(\Omega)|^2 e^{i\Omega \Delta}| \nonumber\\
%N_{1,2} &=& 1 
%\label{J1}
%\end{eqnarray}
%%
%or
%%
%\begin{eqnarray}
%J_K &=& |K| \nonumber\\
%N_{1,2} & = & |{{1} \over {\int d\Omega |G(\Omega) H(\Omega)^*| }}|^2
%\label{J2}
%\end{eqnarray}
%%
%The advantage of Eq.\ref{J1} is that it is simple, trivially reduces to the predictions of standard quantum when $\Delta =0$ and gives intuitive results in the limits of either the detector or source dominating the resolution. Its disadvantage is that it appears contrived and it is hard to see how it might emerge from a more detailed analysis.

%The advantage of Eq.\ref{J2} is that it perhaps seems more natural to make different event functions proportional to their corresponding mode functions, and hence seems less contrived. It also reduces to standard quantum for $\Delta =0$ and behaves qualitatively similarly in the detector and source dominated limits. Its disadvantage is perhaps the rather odd looking normalization condition.

We will assume that the detector has a much sharper intrinsic temporal response than the source. Hong, Ou, Mandel type interference measurements indicate that the intrinsic resolution of silicon APDs is $\le 100 fs$ \cite{HOM}. Under such conditions we can approximate the detector mode function as a constant, $G(k) = 1/\sqrt{2 \pi}$.
\begin{figure}[htb]
\begin{center}
\includegraphics*[width=8.6cm]{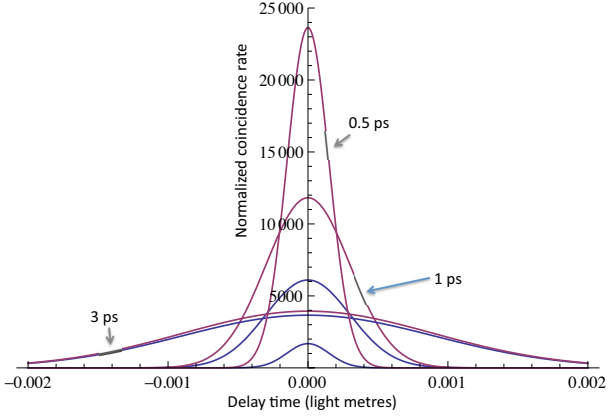}
\caption{Ratio of coincidences to singles as a function of off-set time delay between the detectors in the standard theory (purple) and the event theory (blue). The off-set is chosen such that the maxima lie at $0$. Three different coherence lengths are plotted for a fixed height of 500km.}
\label{fig2}
\end{center}
\end{figure} 
\begin{figure}[htb]
\begin{center}
\includegraphics*[width=8.0cm]{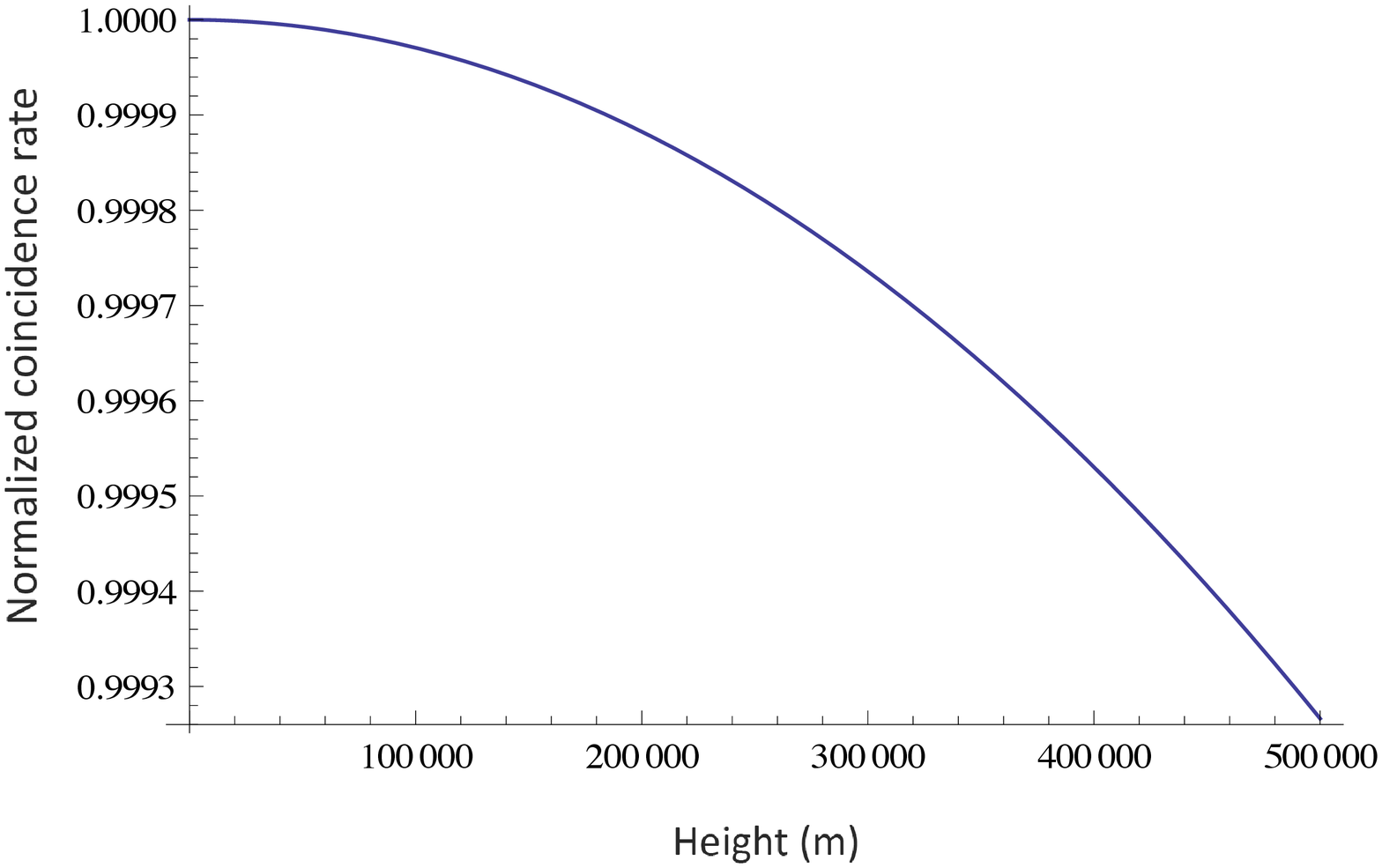}
\caption{Ratio of coincidences to singles as a function of height when the source coherence length is 30ps. Standard quantum mechanics would predict a ratio of 1 for all heights.}
\label{fig3}
\end{center}
\end{figure} 
\begin{figure}[htb]
\begin{center}
\includegraphics*[width=8.0cm]{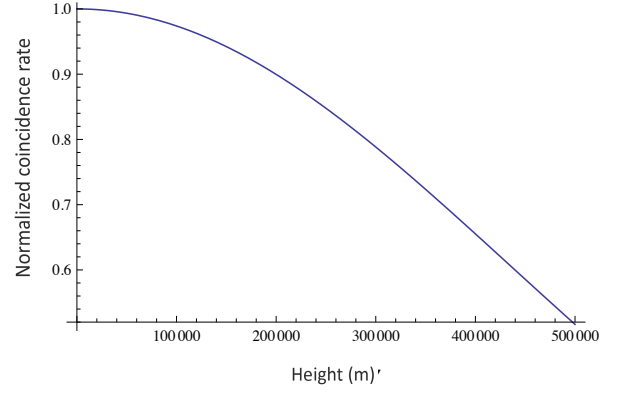}
\caption{Ratio of coincidences to singles as a function of height when the source coherence length is 1ps. Standard quantum mechanics would predict a ratio of 1 for all heights.}
\label{fig4}
\end{center}
\end{figure} 

To estimate the size of the effect we now consider the specific scenario in which the source, mirror, PBS, correlator and detector 1 in Fig.1 are all approximately at height $r_e$, whilst detector 2 is at height $r_e+h$. This corresponds to having the source and detector $1$ on the ground, whilst detector $2$ is on a satellite. A classical channel links the second detector to the correlator on the ground. Eq.\ref{comm} describes the magnitude of $\Delta_t$. Substituting $x_{d1} \approx x_m \approx x_p$ (and maximizing the modal functions by choosing $x_{i1} = x_{i2}$) we have
\begin{eqnarray}
\Delta_t &=& M ln({{x_{d1} x_{i1} x_p^2}\over{x_{d2} x_{i2} x_m^2}})
\nonumber\\
& \approx & M ln({{x_{d1}}\over{x_{d2}}})  \nonumber\\
&\approx& M {{h}\over{r_e}}. 
\label{Del}
\end{eqnarray}
%
%
%experimental apparatus at sea-level where mode 2 is sent vertically a distance $h$ before being reflected back to earth. Assuming the mode function is a Gaussian of the form,
%%
%\begin{equation}
%G(t',x') = {{1}\over{2 \pi d}} e^{-{{(t-t')^2}\over{d_t^2}} - {{(x-x')^2}\over{d_x^2}}}.
%\label{G}
%\end{equation}
%%
%we obtain from Eqs \ref{coin} and \ref{comm}
%%
%\begin{equation}
%C = \chi^2 e^{{{\Delta^2}\over{d_t^2}} + {{\Delta^2}\over{d_x^2}}}.
%\label{C}
%\end{equation}
%
We assume a Gaussian form for the function $H(\Omega)$,
\begin{equation}
H(\Omega) = \sqrt{{{\sqrt{2} d_t}\over{\sqrt{\pi}}}} e^{-(\Omega-\Omega_o)^2 d_t^2}
\label{H}
\end{equation}
In Fig.2 we plot normalized coincidences as a function of the off-set time delay between the detectors (where the off-set has been picked such that zero is the maximum) for the second detector lying at 500km, and for various source coherence lengths. Also shown are the plots expected from standard quantum mechanics (obtained by setting $\Delta_t = 0$). It is seen that as the source coherence lengths become narrower the coincidence counts are suppressed compared to the standard predictions.

It is easier to see what is going on if we assume that the coincidence number is obtained by integrating over the pulse length, i.e. the area under the curves in Fig.2 (this is also a likely scenario for the experiment). We obtain
%Using the assumption of Eq.\ref{J1} we obtain
%%
%\begin{eqnarray}
%C_{\textrm{total}}^{(1)} = \int d t_{d1} \int dt_{d2} \; C & = & |\chi_2|^2 \;\;|\int d\Omega |H(\Omega)|^2 e^{i\Omega \Delta_t}|^2 \nonumber\\
%\label{coin1}
%\end{eqnarray}
%%
%which is essentially the same as in Ref \cite{RMD} except that the function that appears is the source mode function not that of the detector. Using the assumption of Eq.\ref{J2} we get a slightly more complicated expression
%
\begin{eqnarray}
C_{\textrm{total}} = \int d t_{d1} \int dt_{d2} \; C & = & |\chi_2|^2 \;\;|{{\int d\Omega |H(\Omega)| e^{i\Omega \Delta_t}}\over{\int d\Omega |H(\Omega)| }}|^2 \nonumber\\
\label{coin2}
\end{eqnarray}
%For the case of Eq.\ref{J1} we get
%%
%\begin{equation}
%C_{\textrm{total}}^{(1)} = |\chi_{2}|^2 e^{-{{\Delta^2}\over{4 d_{t}^2}}}
%\label{CT1}
%\end{equation}
%%
%For the case of Eq.\ref{J2} we get
and hence we get
\begin{equation}
C_{\textrm{total}}= |\chi_{2}|^2 e^{-{{\Delta_t^2}\over{2 d_{t}^2}}}
\label{CT2}
\end{equation}
The coherence length of current sources suggested for space-based experiments is around $t_c= 30 ps$ \cite{RID12} and hence we set the standard deviation in units of length to $d_t = t_c \times c = 9 \times 10^{-3} m$. Using Eq.\ref{Del}, the mass of earth in units of length, $M = G/c^2 M_{kg} = 4.4 \times 10^{-3} m$ and the radius of earth $r_e = 6.38 \times 10^{6} m$ we find this implies significant decorrelation when $h >10,000 km$ (see Fig.3) - not very practical. In order to get an effect at the height of say the International space-station we need a source with a coherence length $\le 1 ps$ (see Fig.4). 

The coincidence rates in the figures are normalized against the singles rate $|\chi_j|^2$. In the presence of transmission loss it might be better to normalize against the product of the singles rates, thus removing the efficiency, but then $|\chi_1|^2$ and $|\chi_2|^2$ need to be determined independently. In an actual experiment the satellite will be in motion however, the effect of detector $2$'s motion can probably be ignored because the rates are dominated by the source mode function so a Doppler shift on the detector will not significantly affect the result.

\section{The Causal Relationship of Detectors}

In standard quantum mechanics proper and improper mixtures look operationally identical to observers with no information about the way they were created. However, in non-linear extensions of quantum mechanics proper and improper mixtures may become distinguishable. This leads to the so-called preparation problem \cite{CAV12} - when should a particular preparation technique be considered to lead to a proper mixture, and when should it be considered to lead to an improper mixture. The significance of this question is that bad choices can lead to theories which allow instantaneous signalling to occur or other pathologies. 

Two different solutions in the literature, which do not lead to signalling, are due to Bennett et al \cite{BEN09} and Kent \cite{KEN05}. Bennett, et al basically assign an improper mixture to all preparation procedures involving the collapse of a quantum state. This includes not only the collapse of all entangled states but also situations in which states are produced via macroscopic settings, but in accordance with statistics given by a quantum random number generator. In this scenario the only proper mixtures are those produced via macroscopic settings that are determined shot to shot by a deterministic program.

In contrast, Kent assigns a proper mixture to all preparations in which the prepared quantum state lies in the forward light-cone of the preparation outcome. As a result the only improper mixtures in Kent's scheme are those involving entangled states for which the measurement that collapses the state occurs in a region of space-time which is space-like separated from the region in which the non-linear evolution takes place.

Applying the event formalism as so far described, regardless of the space-time relationship of the two detectors, corresponds, in the appropriate limits, to the Bennett, et al solution. Although simple, this solution is not altogether satisfactory \cite{CAV12}. We are thus motivated to make an adjustment to the way the $\Delta$'s are calculated which then corresponds to the Kent solution. 

Consider the geometry of Fig.1. In the previous section we assumed that $t_{d2} > t_{d1}$ and hence that the detectors are space-like separated. We now relax that condition but require that
\begin{eqnarray}
\Delta_t &=& t'_{d1}-\tau_1(t_i, t'_{d1}) - (t_{d2}-\tau_2(t_i, t_{d2})).
\label{comm2}
\end{eqnarray}
where
%\begin{eqnarray}
%t'_{d1} &=& t_{d1} \;\; if x_2 + t_{d2} \ge x_1 + t_{d1} \nonumber \\
%&& {{1}\over{2}}(t_{d1} + t_{d2} +x_2 - x_1 \;\; if \;\; x_2 + t_{d2} \le x_1 + t_{d1} \nonumber \\
%\end{eqnarray}
%
 \begin{eqnarray}
t'_{d1} &=& \bigg\{^{t_{d1} \;\;\;\;\;\;\;\;\;\;\;\;\;\;\;\;\;\;\;\;\;\;\;\;\;\;\;\;   \textrm{if} \;\; x_2 + t_{d2} \ge x_1 + t_{d1}}_{ {{1}\over{2}}(t_{d1} + t_{d2} +x_2 - x_1) \;\; \textrm{if}  \;\; x_2 + t_{d2} < x_1 + t_{d1}} 
\end{eqnarray}
In words, the end point of the evolution for beam $1$, $t'_{d1}$, is either taken to be its detection time, or the point at which beam $1$ enters the forward light cone of the detection point of beam $2$ - whichever comes first. With this assumption the behaviour of the event formalism corresponds, in the appropriate limits, with the Kent solution, and transitions smoothly between those limits. The experimental proposal would be unaffected if the entanglement is distributed directly to the satellite and the ground station. However, if the beam to the satellite was delayed on the ground sufficiently long such that it fell within the forward light cone of the ground detector, then, using Eq.\ref{comm2}, we would predict that the decoherence effect would vanish and the coincidence rates would return to the standard quantum mechanical prediction. This effect could provide a straightforward way to confirm that any decoherence observed in the experiment is due to the physical model of event operators, as opposed to any alternative models or sources of decoherence. In particular, such an effect is certainly not predicted if the decoherence is of a purely semi-classical origin, or if the decoherence were caused by an ordinary coupling to some environmental degrees of freedom, as might be introduced by imperfections in the experimental setup. If a sharp change is seen in the coincidence rates as one adjusts the interval between the detection events, this would provide a `smoking gun' confirmation of the model. On the other hand, if one observes an anomalous decoherence rate, unexplained by other effects but unaffected by the causal relationship of the detectors, this might support the solution of Bennett et al.

\section{Conclusion}

The event operator model is a novel alternative to the standard semi-classical theory of quantum mechanics in curved space-time. It is distinct from most other alternatives because it is based on Deutsch's quantum gravity thought experiment on closed time-like curves, as opposed to Penrose's better known thought experiment of a mass in superposition, the latter being the basis for most other popular non-linear models of decoherence due to gravity. As a consequence, the event operator model makes novel physical predictions in a regime quite different from other models: specifically the distribution of optical entanglement through regions of different curvature, where a general relativistic description is essential. In contrast Penrose type models predict differences when massive objects are put into superposition in Newtonian potentials. The very different regimes of the models leaves open the possibility that they are both limits of some more general model.

%While it is generally difficult to distinguish different possible causes of decoherence, eg. to determine whether some observed loss of coherence is due to the self-gravity of a system (Penrose) versus the number of particles (GRW theory) or some more mundane reason like coupling to an environment, this difficulty does not arise for event operators. In particular, if we see decoherence greater than expected from the standard theory, then we could confirm whether this is caused by the physical mechanism of event operators by moving the detection events until they become time-like separated, and then checking whether the coincidence rates change as a result. An additional check could be provided by varying the height in the gravitational potential as suggested in Ref.\cite{RID}.

Finally, the outcome of such an experiment would have implications for future theoretical work on the topic. Given the possibility of formulating non-pathological non-linear theories, such as by employing the method of Kent\cite{KEN05}, these theories remain an interesting possibility for modelling quantum-gravitational effects. However, such efforts are necessarily contingent on the results of any experiments performed in this new regime, as is our willingness to extrapolate the standard formulation of quantum mechanics into regimes where it might not belong.

{\bf Acknowledgments}: We thank Rupert Ursin and Thomas Scheidl for useful discussions.


\begin{thebibliography}{99}
\bibitem{note} For example, for the boost $v$, the appropriate change of variable is $k' = \sqrt{1-v^2} k$.
\end{thebibliography}
\end{document}